\documentclass{article}
\usepackage{amsmath}
\usepackage{amssymb}
\usepackage{amsfonts}

\begin{document}

\begin{center}
\huge{\bf Supersymmetry in Classical Mechanics}
\bigskip
\bigskip\\
{\Large{E. Deotto, E. Gozzi and D. Mauro}}
\bigskip
\bigskip\\
\large{\it Dipartimento di Fisica Teorica, Universit\`a di Trieste \\
Strada Costiera 11, P.O. Box 586, Trieste, Italy \\
and INFN, Sezione di Trieste}
\end{center}
\bigskip
\bigskip

\noindent In 1931 Koopman and von Neumann [1] proposed an {\it operatorial}
formulation of Classical Mechanics (CM) expanding earlier work of Liouville. Their approach
is basically the following: given a dynamical system with a phase space ${\cal M}$
labelled by coordinates $\varphi^a=(q^i,p^i);\;a=1,\dots,2n;\; i=1,\dots,n$,
with Hamiltonian $H$ and symplectic matrix $\omega^{ab}$, the evolution of a probability 
density $\rho(\varphi)$ can be given either via the Poisson brackets $\{\;,\;\}$ or via the Liouville 
operator:
        \begin{equation}
	\nonumber\tag{1}
	\frac{\partial\rho}{\partial 
t}=\{H,\rho\}=-\hat{L}\rho;\;\;\;\;\widehat{L}=\omega^{ab}\partial_bH\partial_a
	\end{equation}
The evolution via the Liouville operator is basically what is called the operatorial approach to CM. The 
natural question to ask is whether we can associate to the {\it operatorial} formalism of CM a {\it path 
integral} one, like it 
is done in quantum mechanics. The answer is
yes [2]. In fact we can describe the transition probability $P(\varphi^a_{\scriptscriptstyle (2)}t_2|
\varphi^a_{\scriptscriptstyle (1)}t_1)$
of being in configuration $\varphi_{\scriptscriptstyle (2)}$ at time $t_2$ if we were at time $t_1$ in 
configuration 
$\varphi_{\scriptscriptstyle(1)}$ via a functional integral of the form
        \begin{align}
	P(\varphi^a_{\scriptscriptstyle (2)}t_2|\varphi^a_{\scriptscriptstyle(1)}t_1)&=\int {\cal
          D}\varphi^a\,\widetilde{\delta}[
	\varphi^a(t_2)-\varphi^a_{cl}(t_2;\varphi_{\scriptscriptstyle (1)}t_1)]   \nonumber     \\     
	&=\int {\cal D}\varphi^a{\cal D}\lambda_a{\cal D}c^a{\cal D}\bar{c}_a\;e^{i\int
        {\cal \widetilde{L}}dt}  \nonumber\tag{2}
	\end{align}
where $\varphi^a_{cl}(t;\varphi_{\scriptscriptstyle (1)}t_1)$ is the solution of the {\it classical} equations of
motion\newline
$\dot{\varphi}^a=\omega^{ab}\frac{\partial H}{\partial\varphi^b}$ and $\widetilde{\delta}$ is a functional
Dirac delta which gives weight one to the classical paths and zero to the others. In the second line of
(2), via some manipulations [2], we have turned the Dirac $\widetilde{\delta}$ into a more standard looking 
weight 
where
        \begin{equation}
	\nonumber\tag{3}
	{\cal \widetilde{L}}=\lambda_a\dot{\varphi}^a+i\bar{c}_a\dot{c}^a-{\cal \widetilde{H}};
	\;\;\;\;\;\;{\cal \widetilde{H}}=\lambda_a\omega^{ab}\partial_bH+i\bar{c}_a\omega^{ac}\partial_c
	\partial_bHc^b
	\end{equation}
The $\lambda_a,c^a,\bar{c}_a$ are auxiliary variables with $c^a$ and $\bar{c}_a$ of grassmannian character.
The geometrical meaning of these variables
has been studied in [2] and [3]. It is natural at this point to make contact with the operatorial formalism of eq. 
(1). It is easy
to  prove [2] that the first piece of ${\cal \widetilde{H}}$ in (3) is nothing else than the Liouville operator
of (1). To understand the meaning of the full ${\cal \widetilde{H}}$ is important to notice [2] that the $c^a$ 
are nothing
else than the basis $d\varphi^a$ of the forms [4] while the $\bar{c}_a$ are the basis of the vector fields. So 
we can
create a correspondence between forms (tensor fields) and polynomials of $c$ ($\bar{c}$):
        \begin{align}
	F^{\scriptscriptstyle (p)}=\frac{1}{p!}F_{a_1\dots a_p}d\varphi^{a_1}\wedge\cdots\wedge 
d\varphi^{a_p}
	\;\;&\longrightarrow\;\;\widehat{F}^{\scriptscriptstyle (p)}\equiv\frac{1}{p!}F_{a_1\dots 
a_p}c^{a_1}\cdots c^{a_p} \nonumber\\     
        V^{\scriptscriptstyle (p)}=\frac{1}{p!}V^{a_1 \dots a_p}\partial\varphi^{a_1}\wedge\cdots\wedge \partial
        \varphi^{a_p}
	\;\;&\longrightarrow\;\;\widehat{V}^{\scriptscriptstyle (p)}\equiv\frac{1}{p!}V^{a_1 \dots 
a_p}\bar{c}_{a_1}
	\cdots \bar{c}_{a_p}   
	\nonumber\tag{4}
	\end{align}
All the standard operations of the Cartan calculus [4], like the exterior derivative $d$, the interior contraction
$\iota_v$, the symplectic correspondence between forms and Hamiltonian vector fields 
$\alpha=(\alpha^{\sharp})^{\flat}$
can be reproduced via the graded-commutators associated to the path integral (2) in the following way:
        \begin{align}
        &dF^{\scriptscriptstyle (p)}\rightarrow[Q_{\scriptscriptstyle 
BRS}\widehat{F}^{\scriptscriptstyle(p)}],\;\;\;\;\;
        \iota_{\scriptscriptstyle V}F^{\scriptscriptstyle (p)}\rightarrow[\widehat{V},\widehat{F}^
        {\scriptscriptstyle (p)}]\nonumber\\
        &pF^{\scriptscriptstyle (p)}\rightarrow[Q_g,\widehat{F}^{\scriptscriptstyle
        (p)}],\;\;\;\;\;
        V^{\flat}\rightarrow[K,\widehat{V}]\nonumber\\
        &\alpha^{\sharp}\rightarrow[\overline{K},\widehat{\alpha}],\;\;\;\;\;
        (df)^{\sharp}\rightarrow[\overline{Q}_{\scriptscriptstyle BRS},f]\nonumber\tag{5}
        \end{align}
where
        \begin{align}
        &Q_{\scriptscriptstyle BRS} \equiv i c^{a}\lambda_{a};\;\;\;\;\;\;\;\;
        {\overline Q}_{\scriptscriptstyle BRS} \equiv i {\bar c}_{a}\omega^{ab}\lambda_{b}\nonumber \\
        &Q_{g} \equiv c^{a}{\bar c}_{a};\;\;\;\; K \equiv {1\over 2}\omega_{ab}c^{a}c^{b};\;\;\;\;
        {\overline K} \equiv {1\over 2}\omega^{ab}{\bar c}_{a}{\bar c}_{b} \nonumber\tag{6}
        \end{align}
are universally conserved charges under our ${\cal \widetilde{H}}$.
Equipped with this formalism it is then easy to prove [2] that the full ${\cal \widetilde{H}}$
is nothing else than the Lie-derivative ${\cal L}_{(dH)^{\sharp}}=\iota_{(dH)^{\sharp}}d
+d\iota_{(dH)^{\sharp}}$ of the Hamiltonian flow and the correspondence is the following:
        \begin{equation}
	\nonumber\tag{7}
        {\cal L}_{(dH)^{\sharp}}F^{\scriptscriptstyle (p)}\;\;\rightarrow\;\;i[{\cal \widetilde{H}},\widehat{F}^{
        \scriptscriptstyle (p)}]
       	\end{equation}
Beside the five charges in (6) also
$N_{\scriptscriptstyle H}=c^a\partial_aH$ and $\overline{N}_{\scriptscriptstyle 
H}=\bar{c}_a\omega^{ab}\partial_ bH$ 
are conserved under ${\cal \widetilde{H}}$ and as a consequence these other charges are also conserved:
        \begin{equation}
	\nonumber\tag{8}
       	Q_{\scriptscriptstyle H}\equiv Q_{\scriptscriptstyle BRS}-N_{\scriptscriptstyle H};\;\;\;\;\;\;\;
        \overline{Q}_{\scriptscriptstyle H}\equiv\overline{Q}_{\scriptscriptstyle 
BRS}+\overline{N}_{\scriptscriptstyle H}
       	\end{equation}
They are two supersymmetry charges. In fact, while the $Q_{\scriptscriptstyle BRS}$ and 
$\overline{Q}_{\scriptscriptstyle BRS}$ 
anticommute, $Q_{\scriptscriptstyle H}$ and $\overline{Q}_{\scriptscriptstyle H}$ close on ${\cal 
\widetilde{H}}$:
        \begin{equation}
	\nonumber\tag{9}
	[Q_{\scriptscriptstyle H},\overline{Q}_{\scriptscriptstyle H}]=2i{\cal \widetilde{H}}
        \end{equation}
If we enlarge the base space, including two grassmannian partners of time, $\theta$ and $\bar{\theta}$, we 
can put 
together all the 
variables that appear in (3) into the following superfield: $\Phi^a=\varphi^a+\theta
c^a+\bar{\theta}\omega^{ab}\bar{c}_b+i\bar{\theta}\theta\omega^{ab}\lambda_b$. This superfield allows us to 
connect the two
hamiltonians
$H$ and
${\cal
\widetilde{H}}$ via the relation 
${\cal \widetilde{H}}=i\int d\theta d\bar{\theta}H[\Phi]$ and 
to represent the supersymmetry charges as the following operators acting on the superspace 
$(t,\theta,\bar{\theta})$:
$\widehat{Q}_{\scriptscriptstyle H}=-\frac{\partial}{\partial \theta}-\bar{\theta}\frac{\partial}{\partial t},\;
\widehat{\overline{Q}}_{\scriptscriptstyle H}=\frac{\partial}{\partial\bar{\theta}}+\theta\frac{\partial}{\partial t}$. 
This is an N=2 supersymmetry. In fact we could combine the $Q_{\scriptscriptstyle 
BRS},\overline{Q}_{\scriptscriptstyle BRS}, 
N_{\scriptscriptstyle H}, \overline{N}_{\scriptscriptstyle H}$ into the following two charges 
$Q_{\scriptscriptstyle (1)},
Q_{\scriptscriptstyle (2)}$:
        \begin{align}
        &Q_{\scriptscriptstyle (1)}\equiv Q_{\scriptscriptstyle BRS}-\overline{N}_{\scriptscriptstyle H},\;\;\;\;
        Q_{\scriptscriptstyle (2)}\equiv \overline{Q}_{\scriptscriptstyle BRS}+N_{\scriptscriptstyle H} 
\nonumber\\
        &[Q_{\scriptscriptstyle (1)},Q_{\scriptscriptstyle (2)}]=0 \nonumber\tag{10}
        \end{align}
and prove that
        \begin{equation}
        Q_{\scriptscriptstyle (1)}^2=Q^2_{\scriptscriptstyle (2)}=-i{\cal \widetilde{H}} \nonumber\tag{11}
        \end{equation}
As the $Q_{\scriptscriptstyle BRS}$ of (6) is basically the exterior derivative on phase space, it would be 
nice to 
understand the
geometrical meaning also of the susy charges like $Q_{\scriptscriptstyle (1)}$ or $Q_{\scriptscriptstyle 
(2)}$. 
This was done in ref. [5]. The strategy used there was of making local the susy associated to 
$Q_{\scriptscriptstyle (1)}$. The
Lagrangian with this local invariance is
        \begin{equation}
	\nonumber\tag{12}
	\widetilde{\mathcal L}_{\scriptscriptstyle EQ} := \widetilde{\mathcal L} + 
	\alpha(t)Q_{\scriptscriptstyle (1)} + g(t)\widetilde{\mathcal H}  
	\end{equation}
where $\alpha(t)$ and $g(t)$ are gauge fields. The physical-state conditions associated to this gauge 
invariance
turns out to be:
        \begin{align}	
	&\widetilde{\mathcal H}|\mbox{phys}\rangle = 0\nonumber\\
	&Q_{\scriptscriptstyle (1)} |\mbox{phys}\rangle = 0\nonumber \\
	&\Pi_{\scriptscriptstyle g}|\mbox{phys}\rangle = 0 \nonumber\\
	&\Pi_{\scriptscriptstyle \alpha}|\mbox{phys}\rangle = 0 \nonumber\tag{13}
	\end{align}
where $\Pi_g$ and $\Pi_{\alpha}$ are the momenta associated to the gauge variables $\alpha,g$. Using the 
correspondence
(4)-(7) it is easy to translate (13) into a differential-geometric language and prove that the states selected 
by (13) 
are in one-to-one 
correspondence with the states of the so-called equivariant cohomology [6] with respect to the 
Hamiltonian vector 
field. The equivariant cohomology w.r.t. a vector field V is defined as the set of forms $|\rho\rangle$ which
satisfy the following conditions:
        \begin{align}
	&(d - \iota_{\scriptscriptstyle V})|\rho\rangle  = 0 \nonumber\\
	&{\cal L}_{\scriptscriptstyle V}|\rho\rangle = 0 \nonumber\\
	&|\rho\rangle \neq (d - \iota_{\scriptscriptstyle V})|\chi\rangle \nonumber\\
	&{\cal L}_{\scriptscriptstyle V}|\chi\rangle = 0 \nonumber\tag{14}
	\end{align}
This is the geometrical light we could throw on the susy charge $Q_{\scriptscriptstyle (1)}$. Our universal 
symmetries, 
besides having a nice geometrical interpretation, should also have a {\it dynamical} meaning. 
This is the case for the susy invariance
which seems to have some interplay with the concept of ergodicity [5]-[7]. We will not expand on it
here but turn to another aspect of this susy.

Supersymmetry has found its most important applications in field theory where it has produced theories 
which have a better
ultraviolet behaviour than non supersymmetric ones. With that in mind in ref. [9] an attempt was made 
to build the analog
of ${\cal \widetilde{H}}$ and ${\cal \widetilde{L}}$ of eq. (3) also for field theory.
Starting for example from the Hamiltonian of a $\varphi^4$ theory 
${\cal{H}}_{\varphi^4}=\int 
d^3x\;\{\frac{1}{2}\Pi_{\varphi}^2+\frac{1}{2}(\partial_k\varphi)^2+\frac{1}{2}m^2\varphi^2+
\frac{1}{4!}g\varphi^4\}$
the associated ${\cal \widetilde{L}}$ is
        \begin{equation}
	\nonumber\tag{15}
	{\cal \widetilde{L}}_{\varphi^{4}}=\int\{\Lambda_a(\dot{\xi}^a-
        \omega^{ab}\delta_bH_{\varphi^4})+i\overline{\Gamma}_a(\partial_t\delta^a_b-
        \omega^{ac}\delta_c\delta_bH_{\varphi^4})\Gamma^b\}d^3x
	\end{equation}
where $\xi^a$ are made of $(\varphi,\Pi_{\varphi})$ and $\Lambda_a(\vec{x},t), \Gamma^a(\vec{x},t),
\overline{\Gamma}_a(\vec{x},t)$ are the fields analogous to the point-particle variables 
$\lambda_a,c^a,\bar{c}_a$
but, differently from these ones, they do not depend only on $t$ but also on $\vec{x}$. Like for the point 
particle,
${\cal \widetilde{L}}_{\varphi^{4}}$ has an N=2 susy whose charges are:
        \begin{align}
	&Q_{\scriptscriptstyle H}=\int d^3x (i\Gamma^a\Lambda_a-\Gamma^a\delta_a H_{\varphi^4}) 
\nonumber\\
	&\overline{Q}_{\scriptscriptstyle H}=\int 
d^3x(i\overline{\Gamma}_a\omega^{ab}\Lambda_b+\overline
	{\Gamma}_a\omega^{ab}
	\delta_bH_{\varphi^4}) \nonumber\tag{16}
	\end{align}
The same construction can be done for any field theory and even for gauge theories. For example the 
standard BFV 
Hamiltonian for Yang-Mills theories is [8]:
        \begin{align}
	&H_{\scriptscriptstyle BFV}=
        \int d^3x \biggl\{\frac{1}{2}\pi^k_a\pi^{a}_k+\frac{1}{4}F_a^{ij}F^{a}_{ij}
        +\pi_a\partial^kA^{a}_k-
        \lambda^{a}\partial_k\pi^k_a+\lambda^{a}C^b_{\;ac}
        \pi^k_bA^c_k \nonumber\\
        &+i\overline{P}_aP^{a}-\lambda^{a}\overline{P}_bC^b_{\;ac}C^c
        -i\overline{C}_a\partial^k(\partial_kC^{a}+C^{a}_{\;bc}A^c_kC^b)\biggr\} \nonumber\tag{17}
        \end{align}
where $A^k_a$ are the gauge fields, $\pi_k^a$ are their conjugate momenta, $F^a_{ij}$ is the antysimmetric
tensor, $\lambda^a$ is a Lagrange multiplier and $\pi_a$ its conjugate momentum, $(-iP^a,C^a)$ are the 
BFV
ghosts of the theory and $(i\overline{C}_a,\overline{P}_a)$ the BFV anti-ghosts. If we indicate with $\xi^A$ 
all the fields
of the theory above, including the BFV ghosts, we have that the associated ${\cal \widetilde{H}}$ is [9]:
        \begin{equation}
        \widetilde{\cal H}_{\scriptscriptstyle BFV}=\int d^3x\{\Lambda_A 
\omega^{AB}\vec{\delta}_BH_{\scriptscriptstyle
        BFV}(\xi)
        +i\bar{\Gamma}_A\omega^{AC}\stackrel{\rightarrow}{\delta}
        _CH_{\scriptscriptstyle BFV}(\xi)\stackrel{\leftarrow}{\delta}_B\Gamma^B\} \nonumber\tag{18}
        \end{equation}
where $\Lambda_A, \Gamma^ A, \overline{\Gamma}_A$ are auxiliary fields. 
$\Lambda$ has the same grassmannian parity
as the field $\xi$ to which it refers, while 
$\Gamma$ and $\overline{\Gamma}$ have opposite grassmannian parity.
It is easy to show that there are conserved BRS and anti-BRS charges of the form: 
$Q=i\int d^3x \Gamma^A\Lambda_A$ and 
$\overline{Q}=-i\int d^3x \Lambda_A\omega^{AB}\overline{\Gamma}_B$. The Hamiltonian of eq. (18) can be
written as a pure BRS variation in the following way: 
        \begin{equation}
        \widetilde{\cal H}_{\scriptscriptstyle BFV}=
        -i[Q,[\overline{Q},H_{\scriptscriptstyle BFV}]] \nonumber\tag{19}
        \end{equation}
This is not a surprise because it is a property of any Lie-derivative. It makes this $\widetilde{\cal
H}_{\scriptscriptstyle BFV}$ strongly similar to the hamiltonians of Topological Field Theories [8]. 
The susy charges $Q_{\scriptscriptstyle H}$ and $\overline{Q}_{\scriptscriptstyle H}$ have the same form
as the ones of the $\varphi^4$ theory, except for
the presence of some grading factors. They are:
        \begin{align}
        &Q_{\scriptscriptstyle H}=\int d^3x (i\Gamma^A\Lambda_A- (-)^{[\xi^A]}\Gamma^A
        \stackrel{\rightarrow}{\delta}_A H_{\scriptscriptstyle BFV})\nonumber\\
        &\overline{Q}_{\scriptscriptstyle H}=\int d^3x(-i\Lambda_A\omega^{AB}\overline{\Gamma}_B
        +\overline{\Gamma}_A\omega^{AB}\stackrel{\rightarrow}{\delta}_BH_{\scriptscriptstyle 
BFV})\nonumber\tag{20}
        \end{align}
where we indicate with $[\xi^A]$ the grassmannian parity of the field $\xi^A$. Their
anticommutator produces the Hamiltonian (18): $[Q_{\scriptscriptstyle H},\overline{Q}_{\scriptscriptstyle 
H}]=2i\int d^3x 
{\cal \widetilde H}_{\scriptscriptstyle BFV}$.  The shortcoming of all this
is that we have obtained a {\it non-relativistic} susy even from a {\it relativistic} field theory. 
We feel anyhow that it should be possible
to get a relativistic one. The strategy should be to start {\it not} from the Hamiltonian
formalism but from an explicitly Lorentz covariant one like the DeDonder-Weyl approach [10]. 
The Hamiltonian formalism gives a special role to time and spoils the manifest Lorentz covariance. This 
special role 
of time is what produces a non-relativistic susy in our formalism. If we succeed in getting a relativistic 
susy with our mechanism we can say that somehow {\it susy is everywhere}, even associated to a non-susy 
theory like
a $\varphi^4$-theory. We haven't seen this susy before because we haven't considered all the other 
geometrical fields (forms and vector fields) which are naturally associated with the basic fields
$\varphi$. The susy appeared only when we did things in a coordinate 
indipendent fashion as the Lie-derivative does. 

\bigskip
\noindent{\bf BIBLIOGRAPHY}

\noindent [1] B.O. Koopman Proc.Nat.Acad.Sci. USA {\bf 17} (1931) 315; \newline J.von Neumann
Ann.Math. {\bf 33} (1932) 587; ibid. {\bf 33} (1932) 789\newline 
[2] E. Gozzi, M. Reuter and W.D. Thacker Phys.Rev.D {\bf 40} (1989) 3363\newline
[3] E. Gozzi and M. Regini Phys.Rev.D  {\bf 62} (2000) 067702 [hep-th/9903136];\newline 
E. Gozzi and D. Mauro Jour.Math.Phys {\bf 41} (2000) 1916 [hep-th/9907065]\newline
[4] R. Abraham and J. Marsden {\it "Foundations of Mechanics"}, Benjamin 1978\newline
[5] E. Deotto and E. Gozzi hep-th/0012177\newline
[6] H. Cartan {\it "Colloque de Topologie"} (Espace Fibres), CBRM 15.71 1950\newline
[7] E. Gozzi and M. Reuter Phys.Lett. {\bf 233B} (1989) 383; Chaos, Solitons and Fractals {\bf 2} (1992) 
441;\;
 V.I. Arnold and A. Avez {\it "Ergodic Problems of Classical Mechanics}, W.A. Benjamin Inc. 1968\newline
[8] M. Henneaux Phys. Rep. {\bf 126} (1985) 1\newline
E. Gozzi e M. Reuter Phys. Lett. B {\bf 240} (1990) 137\newline
[9] P. Carta Master Thesis, Cagliari University 1994; 
\newline D. Mauro Master Thesis, Trieste University 1999\newline
[10] H.A. Kastrup Phys. Rep. {\bf 101} (1983) 1

\end{document}